\documentclass[prd,preprint,epsfig]{revtex4}

\usepackage{graphicx}
\usepackage{dcolumn}
\usepackage{bm}

\newcommand{\bq}{\begin{equation}}
\newcommand{\eq}{\end{equation}}
\newcommand{\bqa}{\begin{eqnarray}}
\newcommand{\eqa}{\end{eqnarray}}
\newcommand{\nn}{\nonumber \\}

\begin{document}
\draft
\title{Emergence of gravity from interacting simplices}

\author{
Sung-Sik Lee
}
\address{
Department of Physics $\&$ Astronomy, McMaster University,
Hamilton, Ontario L8S 4M1, Canada
}

\begin{abstract}
We consider a statistical model of interacting 4-simplices fluctuating in an $N$-dimensional target space.
We argue that a gravitational theory may arise as a low energy effective theory in a strongly interacting phase 
where the simplices form clusters with an emergent space and time with the Euclidean signature.
In the large $N$ limit, two possible phases are discussed, that is,
 `gravitational Coulomb phase' 
and  `gravitational Higgs phase'.
\end{abstract}
\date{\today}

\maketitle

\noindent

\section{Introduction}

Earlier, Sakharov pointed out that the Einstein's gravitational action can be generated solely 
from quantum fluctuations of a matter field defined on a curved Riemannian manifold\cite{SAKHAROV}.
In this proposal, the bare gravitational coupling is infinite and the metric does not have any microscopic dynamics.
However, the curvature term is generated as high energy modes of the matter field are integrated out.
Then the graviton becomes a propagating mode at low energies.

One can view the Sakharov's observation from a different perspective.
Since the metric does not have any microscopic dynamics, one can regard it as an auxiliary field.
If one integrates out the auxiliary metric field first, one obtains an action for the matter field alone. 
The resulting action for the matter field is local because the integrated metric field is not a microscopic propagating mode.
This action can, in principle, be realized as a local microscopic model which has only the matter field as fundamental degrees of freedom\cite{NONLOCAL}.
Then, whatever arises as a low energy excitation of the model should be regarded as an excitation made of the microscopic degree of freedom.
In particular, if spin-2 graviton arises as a low energy excitation through the Sakharov's mechanism, the graviton is nothing but a collective excitation of the matter field.
In this sense, the gravity can emerge from the microscopic model which does not have metric as a fundamental degree of freedom.

This mechanism for emergent gravity applies to emergent gauge theory too.
The possibility of obtaining  emergent gauge theories in this way has been extensively studied
through so called `slave-particle' theories for condensed matter systems\cite{ANDERSON,BASKARAN,AFFLECK,IOFFE,PLEE,SENTHIL2000}.
In this approach, there is no microscopic degree of freedom which is to become a gauge field at low energies.
The microscopic degrees of freedom consist of only `gauge neutral' particles.
When there are strong interactions between those particles, dynamical constraints force new degrees of freedom to emerge at low energies.
Usually, microscopic particles form a composite particle which becomes a low energy excitation.
However, the opposite can also happen, that is, the microscopic field is forced to be written as a composite field of low energy fields which are called slave-particle (or parton) fields.
The effective theory for the slave-particles is highly nonlinear due to the composite nature of the original microscopic field. 
The nonlinear action can be cast into a quadratic form through a Hubbard-Stratonovich transformation where the Hubbard-Stratonovich field plays the role of a gauge field.
Because one started with the gauge neutral particles, 
the slave-particles are not `free' but they are
coupled to the gauge field with an infinite gauge coupling.
Here the gauge field is an auxiliary field which ensures that
the slave-particles do not escape from the original particle.
If one integrates out the auxiliary gauge field, one obtains the original theory back, which is difficult to analyze.
Instead, if we integrate out high energy modes of the matter fields, the gauge coupling gets renormalized to a finite value due to loop corrections.
If the gauge coupling flows to a sufficiently small value at low energies, a deconfinement phase can arise.
The gauge coupling is marginal in 3+1D and it is relatively easy to stabilize the deconfinement phase in dimensions higher than 3+1D.
In a lower dimension, e.g. 2+1D, massless fundamental matters are usually required to stabilize the deconfinement phase with a continuous gauge group\cite{SENTHIL04,HERMELE,LEE2008}.
In the deconfinement phase, the slave-particles are effectively `liberated' and a massless photon emerges instead of the confining gauge field.
The emergent photon is not a fundamental excitation but a collective excitation of the microscopic degree of freedom.
This phenomenon dubbed as {\it fractionalization} is possible although the microscopic slave-particles
are always `confined' within the gauge neutral particle\cite{LEE}.

In this paper, we will apply the strategy analogous to slave-particle theories to emergent gravity\cite{OTHER}.
We will consider a local model whose fundamental degrees of freedom are simplices fluctuating in a target space.
The simplices have no reference to a space-time hence they are diffeomorphism invariant degrees of freedom (like the gauge neutral particles in slave-particle gauge theories).
When there is no interaction between the simplices, they form a gaseous phase in the target space.
In this phase, there is neither an extended space-time nor a metric.
When there is a strong interaction, the simplices are subject to a dynamical constraint
which forces them to form clusters (or discrete membranes) in the target space.
The vertices of the discrete membrane become the new low energy degrees of freedom
and the way the vertices are connected to each other in the target space defines
an emergent space and time with the Euclidean signature.
As is the case for slave-particle theories, the theory for the membrane is highly nonlinear 
and the action can be transformed into a quadratic form by introducing an auxiliary field.
In this case the auxiliary field plays the role of a metric.
Actually, the nonlinear action for the membrane
can be viewed as a theory which is obtained after integrating out the auxiliary metric field 
in a model which has both the membrane modes and the auxiliary metric field.
In this sense, the membrane theory is closely related to a gravitational theory with an infinite coupling.
In the low energy limit, the auxiliary metric field acquires the curvature term through the Sakharov's mechanism
and becomes a propagating mode.

One may ask what the point of emergent gravity is if one obtains a gravitational theory at low energies
starting from a microscopic model which can be obtained from another model with the auxiliary metric field.
The point is that the local action which is obtained by integrating out the auxiliary metric field
can, in principle, arise from a variety of physical systems.
The hope is that a better understanding of the microscopic model 
can help identify physical systems which may show emergent gravity.
The situation is clearer for emergent gauge theories.
If one integrates out an auxiliary gauge field in a gauge theory with an infinite gauge coupling, one obtains a local theory of  gauge neutral matter fields.
Such a theory of the gauge neutral particles can arise in many condensed matter systems, i.e., quantum magnets which seemingly have nothing to do with gauge theories\cite{PLEE}.
In priori, it is not clear whether such systems have a gauge boson as a low energy excitation.
It is a complicated dynamical issue.
If a gauge boson indeed arises as a low energy excitation, it is purely a collective excitation of the microscopic degree of freedom, i.e., spins in a magnet.
Although searches for concrete experimental realizations are still underway\cite{SHIMIZU1,HELTON}, gauge theory can, in principle, be emergent.

\section{Emergent gauge theory}
Before we discuss about emergent gravity, 
we review a simple model which exhibits emergent gauge theory.
We consider the following model\cite{LEE},
\bqa
S_0 & = & -K \sum_{<i,j>} \sum_{a<b} \cos( \theta^{ab}_i - \theta^{ab}_j ) \nn
&& -  K_3 \sum_i \sum_{a,b,c} \cos ( \theta^{ab}_i + \theta^{bc}_i + \theta^{ca}_i ).
\label{b1_K3}
\eqa
Here $\theta^{ab}_i$ are compact XY-variables where $i$ is 
a site index in the 4 dimensional hyper-cubic lattice, and
$a,b =1,2,...,N$ are flavor indices.
$<i,j>$ refers to nearest neighbor sites in the hyper-cubic lattice.
$\theta^{ab}_i$ are the U(1) phases of a Hermitian matrix
and they satisfy the constraint $\theta^{ab}_i = - \theta^{ba}_i$.
There are $\frac{N (N-1)}{2}$ independent variables at each site.
Although the above model can be derived as an effective theory of a multi-band exciton bose condensate\cite{LEE},
here we regard the model as our starting point.
The 4-dimensional hyper-cubic lattice is made of three dimensional spatial lattice 
and one imaginary time.
The first term in Eq. (\ref{b1_K3}) is the lattice form of the kinetic energy of the relativistic bosons
and the second term describes interactions between the bosons.
In the weak coupling limit $K_3 << 1$, the model  simply describes $\frac{N (N-1)}{2}$ copies of the 4D XY model
which has the disordered phase and the bose condensed phase depending on the value of $K$.
In the disordered phase, all excitations are gapped.
In the bose condensed phase, there are $\frac{N (N-1)}{2}$ gapless Goldstone modes.
In the strong coupling limit $K_3 >> 1$, 
the strong interaction imposes a set of dynamics constraints,
\bqa
\theta^{ab}_i + \theta^{bc}_i + \theta^{ca}_i = 0
\eqa
for all possible combinations of $a,b,c$.
This constraint is solved by decomposing the microscopic field into two parton fields as
\bqa
\theta^{ab}_i = \phi^a_i - \phi^b_i
\label{sl}
\eqa
and $\phi^a_i$ becomes the low energy degree of freedom.
The effective theory for the low energy fields becomes
\bqa
S & = & -\frac{K}{2} \sum_{<i,j>} \sum_{a<b}  \left[ e^{ i ( \phi^{a}_i - \phi^{a}_j )} e^{ - i ( \phi^{b}_i - \phi^{b}_j )} + c.c. \right].
\eqa
Since the action is nonlinear with respect to the slave-boson field $e^{i\phi^a}$, 
it is not easy to figure out low energy excitations of this model.
To make the action quadratic for the boson field, we 
decouple the quartic term through the Hubbard-Stratonovich transformation\cite{LEE},
\bqa
S^{'} & = &   \frac{K (N-1)}{2} \sum_a \sum_{<i,j>} 
\Bigl[   |\eta_{ij}|^2 - |\eta_{ij}|  e^{ i ( \phi^{a}_i - \phi^{a}_j - a_{ij}) } - c.c.  \Bigr],
\label{boson_action}
\eqa
where $\eta_{ij} = |\eta_{ij}| e^{i a_{ij}}$ is the complex auxiliary field defined on every link of the lattice.
The phase of the complex auxiliary field becomes a gauge field.
This is the compact U(1) gauge theory coupled with the N bosons.
Since $a_{ij}$ has no dynamics, the bare gauge coupling is infinite.
However, the gauge coupling is renormalized to a finite value at low energies
as high energy modes of $\phi^a$ are integrated out.
For a small $t = K(N-1)$, the slave bosons are gapped.
If we integrate out the slave bosons completely, the gauge coupling is renormalized to $g^2 \sim \frac{1}{t^4 N}$.
When $N>>1$, the renormalized gauge coupling can be made smaller than $1$ and 
the deconfinement phase can be stabilized.
Although the microscopic slave-particles are always confined within the original bosons $\theta^{ab}$, 
they can effectively be deconfined by propagating in the medium of other slave-particles
as it keeps changing its partners\cite{LEE}.
Then a gapless emergent gauge boson arises as a low energy excitation. 
The emergent gauge boson (or photon) is nothing but a collective excitation of the original XY-bosons $\theta^{ab}$.
This is because the microscopic model in Eq. (\ref{b1_K3}) has only the XY-bosons as its fundamental degrees of freedom.
The emergent photon can be probed through a correlation function of a composite operator of the original bosons\cite{LEE}.

In the following, we will construct a model which shows an emergent gravity 
in the similar fashion as the gauge theory emerges from the model  (\ref{b1_K3}).

\section{Emergent gravity}

\begin{figure}
        \includegraphics[height=7cm,width=7cm]{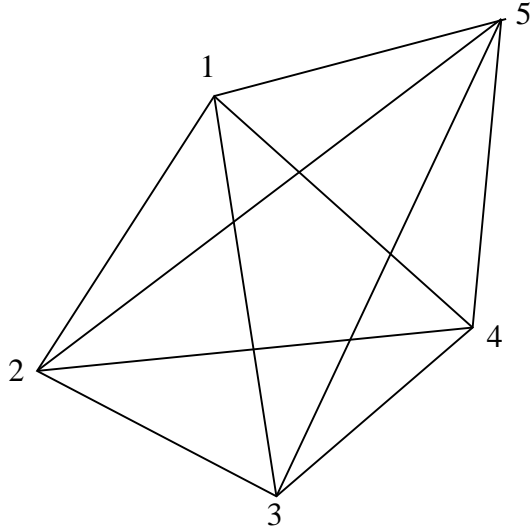}
\caption{
A 4-simplex in the N-dimensional target space.
}
\label{fig:4simplex}
\end{figure}

\begin{figure}
        \includegraphics[height=7cm,width=14cm]{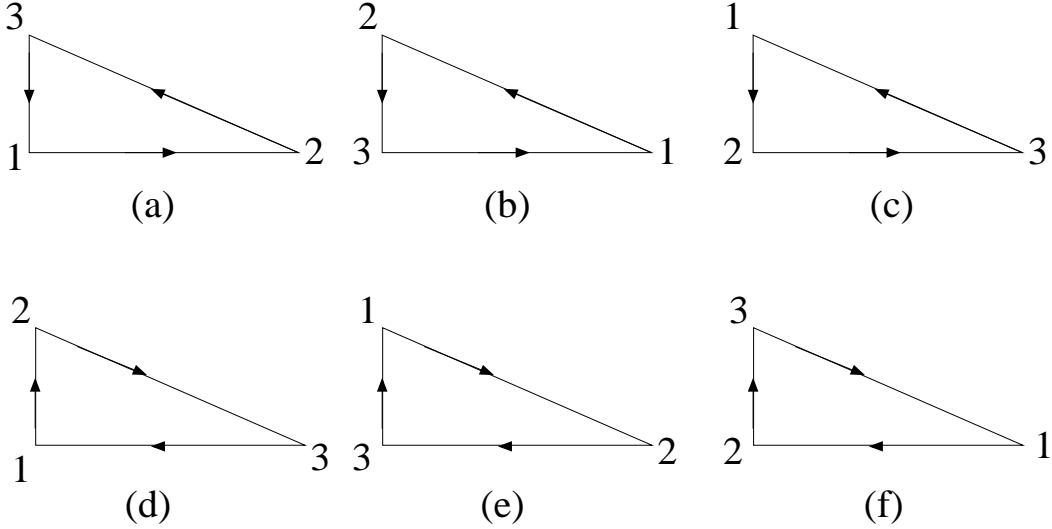}
\caption{
Oriented 2-simplex (triangle) with indistinguishable vertices.
The arrows represent the orientation of faces.
For a fixed set of coordinates of vertices, there are two different configurations.
(a),(b),(c) represent one configuration with one orientation
and (d),(e),(f), the other configuration with the opposite orientation. 
}
\label{fig:permu}
\end{figure}

We consider orientable 4-simplices fluctuating in an $N$-dimensional target space, 
$\{ \phi^A \}$ with $A=1,2,...,N$.
Here we assume that $N >> 1$.
The fundamental degrees of freedom are shapes and positions of simplices
which are determined by coordinates of vertices in the target space.
Each 4-simplex has five vertices as is shown in Fig. \ref{fig:4simplex} 
and the coordinate of each vertex is given by 
an $N$-component vector, $\phi^A_{sv}$, 
where $s=1,2,..,M$ is an index for simplices and 
$v=1,2,..,5$ is an index for vertices.
The partition function is written as
\bqa
Z & = & \sum_{M=0}^\infty \frac{1}{M!(5!/2)^M} \int \Pi_{s=1}^M \Pi_{v=1}^5 \Pi_{A=1}^N d \phi^{A}_{sv}~~ e^{- S^M[ \phi^A_{sv}] }.
\label{p}
\eqa
Here we are using the grand canonical ensemble where
the number of simplices, $M$ are allowed to fluctuate.
The factor $M!$ is due to the fact that simplices are indistinguishable.
The factor $5!/2$ is the number of permutations of vertices which do not change the orientation of a simplex.
It takes into account of the fact that vertices of a simplex with a fixed orientation are indistinguishable.
In other words, an even permutation of $v=1,2,...,5$ does not create a new simplex but
an odd permutation does create a new simplex with the opposite orientation.
This is illustrated in Fig. \ref{fig:permu} for a two simplex.
The generalization to four simplex is straightforward.

The action for simplices consists of two parts, $S^M=S^M_1 + S^M_2$.
The first term $S_1^M$ is an `one-body' term,
\bqa
S_1^M & = & \sum_{s=1}^M \left[ f(V_s ) - \mu \right].
\label{action1}
\eqa
Here $V_s$ is the volume of simplex $s$ in the target space
which is obtained by using the Euclidean metric, and $\mu$ is the chemical potential.
$f(V_s)$ controls volumes of simplices and $\mu$ controls the number of simplices.
$f(x)$ is defined as
\bqa
f(x) & = & - \ln \int d {\bf E} 
\exp \Bigl[  -\sqrt{g} \Bigl( 
1 
+ {\mbox tr} \sum_{n=1}^\infty \alpha_n x^{n/2} ({\bf G}^{-1})^n  
\Bigr)  \Bigr].
\label{f}
\eqa
Here ${\bf E}$ is a $4 \times 4$ real matrix and the integration is over all $16$ components of ${\bf E}$.
${\bf G} = {\bf E} {\bf E}^{T}$ is a symmetric matrix,
${\bf G}^{-1}$, the inverse of ${\bf G}$ and
$g$, the determinant of  ${\bf G}$.
$g \geq 0$ because eigenvalues of ${\bf G}$ are non-negative.
$\alpha_n$'s are constants.
${\bf E}$ is an auxiliary field which is used to define the action for the simplices.
We will see that the auxiliary field becomes the tetrad field 
when simplices are strongly interacting with each other.

Unfortunately, it is difficult to perform the integration over the auxiliary field to write down $f(x)$ in a closed form.
Here we provide only the asymptotic form of $f(x)$ in the small $x$ limit.
The real matrix ${\bf E}$ can be decomposed as ${\bf E}= {\bf O}_1^T {\bf D} {\bf O}_2$,
where ${\bf D}$ is a diagonal matrix with positive elements $\{ \lambda_1, \lambda_2, \lambda_3, \lambda_4 \}$
and ${\bf O}_1$, ${\bf O}_2$ are orthonormal matrices. 
Since the integrand in the right hand side of Eq. (\ref{f}) is independent of ${\bf O}_1$, ${\bf O}_2$,
the integrations over the orthonormal matrices simply result in the measure
$\Pi_{i>j} | \lambda_i^2 - \lambda_j^2 |$ upto a multiplicative constant.
Rescaling the diagonal elements as $\lambda_i \rightarrow x^{1/4} \lambda_i$,
we obtain 
\bqa
f(x) & = & - \ln \Bigl\{ x^4 \int d \lambda_1 d \lambda_2 d \lambda _3 d \lambda_4 ~~\Pi_{i>j} | \lambda_i^2 - \lambda_j^2 |   \nn
&&
\exp \Bigl[  - x \lambda_1 \lambda_2 \lambda_3 \lambda_4  \Bigl( 
1 + \sum_{n=1}^\infty \alpha_n \left( 1/\lambda_1^2 + 1/\lambda_2^2 + 1/\lambda_3^2 + 1/\lambda_4^2 \right)^n  
 \Bigr)  \Bigr] \Bigr\},
\label{f2}
\eqa
where we ignored an additive constant.
Here we assume that $\alpha_n$'s are mostly positive that the term  
$\sum_{n=1}^\infty \alpha_n \left( 1/\lambda_1^2 + 1/\lambda_2^2 + 1/\lambda_3^2 + 1/\lambda_4^2 \right)^n $
in the exponent becomes positive infinite as $\lambda_i \rightarrow 0$.
It provides a lower cut-off for $\lambda_i$ but it is negligible for large $\lambda_i$.
Therefore, we can approximate $f(x)$ as 
\bqa
f(x) & \approx & - \ln \Bigl\{ x^4 \int_{ |\lambda_i|>a} d \lambda_1 d \lambda_2 d \lambda _3 d \lambda_4 ~~\Pi_{i>j} | \lambda_i^2 - \lambda_j^2 |  \exp \Bigl[  - x \lambda_1 \lambda_2 \lambda_3 \lambda_4   \Bigr] \Bigr\},
\label{f3}
\eqa
where $a>0$ is an effective lower cut-off which would be determined from $\alpha_n$'s.
This integration can be performed. 
In the small $x$ limit, we obtain
\bqa
f(x) & = &  \ln x^3 + ...,
\label{f4}
\eqa
where $...$ represents terms which depend on $a$ (hence depend on $\alpha_n$) and are finite in the $x \rightarrow 0$ limit.
For $x \rightarrow 0$, the one-body potential becomes negative infinite.
Because of this, the partition function may have divergent contributions from configurations with vanishingly small simplices.
The divergence would imply that all simplices collapse to point-like objects in the target space
and there will be no low energy degree of freedom except for the center of mass motion of the simplices.
However, the divergence is absent for a large enough $N$, where $N$ is the dimension of the target space.
The integration over the vertices of simplices in the $N$-dimensional target space in Eq. (\ref{p}) induces a measure of the order of $V_s^k$ with $k \sim N$.
This induced measure vanishes as the volume goes to zero and it can cancel the divergent contributions from small world volumes.
Therefore, we expect to have a non-trivial simplicial dynamics for a large enough $N$.

A few comments are in order regarding the one-body action.
First, the auxiliary field ${\bf E}$ is not a fundamental degree of freedom of the microscopic model.
The only microscopic degrees of freedom are shapes and positions of simplices.
In principle, we could have simulated the action given by Eqs. (\ref{action1}) and (\ref{f4}) in a computer
to study the dynamics of simplices without introducing the auxiliary field.
Therefore, any low energy excitation of the model should be regarded as a part of fluctuations of the simplices.
Second, as it will become clear in the followings, 
the form of the action in Eq. (\ref{f}) is constructed so that
the low energy effective theory is described by a gravitational theory.
However, we expect that a large class of function which has the leading logarithmic singularity as in Eq. (\ref{f4})
can be written in the form of Eq. (\ref{f}) for some $\alpha_n$'s.

\begin{figure}
        \includegraphics[height=5cm,width=10cm]{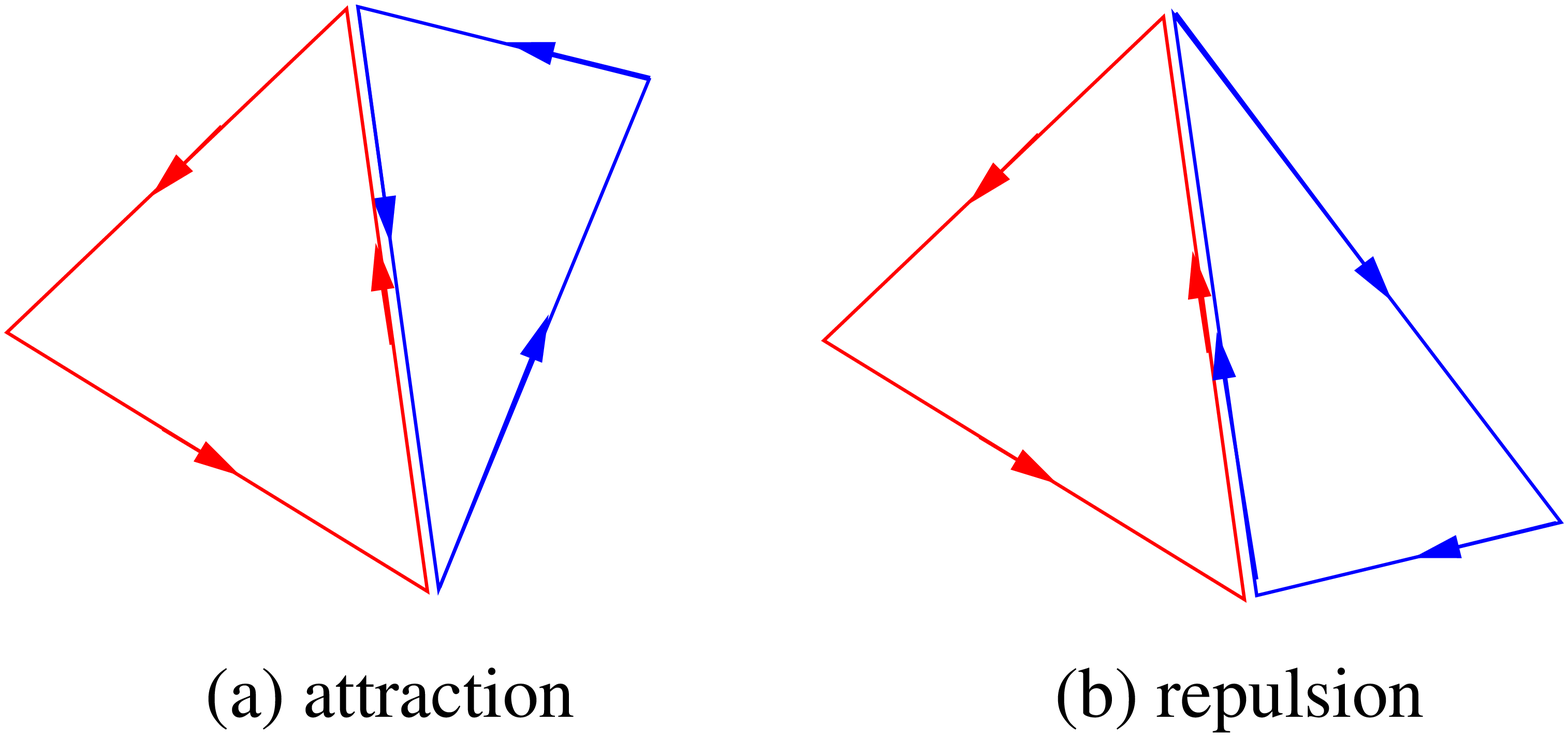}
\caption{
The interaction between 2-simplices.
If a face of a simplex touch a face of another simplex with the same (opposite) orientation, 
the energy increases (decreases). 
The interaction between 4-simplices can be generalized in the same way.
}
\label{fig:int}
\end{figure}

The second term in the action describes a pairwise interaction between simplices,
\bqa
S^M_2 & = & U \sum_{s>s^{'}} \sum_{f,f^{'}=1}^5 \sum_P (-1)^P \Pi_{k=1}^4 \Pi_{A=1}^N  
\delta( \phi^A_{sv(f,k)} - \phi^A_{s^{'}v(f^{'},P(k))} ).
\label{action2}
\eqa
Here $U \geq 0$ is the strength of the interaction.
$f=1,2,..,5$ labels five faces (3-simplices) of each 4-simplex
and $k$ is an index for four vertices of each face.
$v(f,k)$ is an integer function which specifies vertices of a 4-simplex
as a function of $f$ and $k$, as is shown in Table \ref{t1}.
$P$ is a permutation of $1,2,3,4$ with $(-1)^P = 1 (-1)$ for an even (odd) permutation.
$S^M_2$ describes an interaction between two 4-simplices, where
the energy is increased (lowered) when faces of 
two 4-simplices with a same orientation (opposite orientations) contact with each other.
This is illustrated in Fig. \ref{fig:int}.

\begin{table}
 \begin{center}
 \begin{tabular}{|c||c|c|c|c|}
\hline
~~f $\backslash$ k ~~& ~~1~~ & ~~2~~ & ~~3~~ & ~~4~~ \\
\hline
\hline 
1 &   2& 3& 4& 5 \\
\hline
2 &   3& 1& 4& 5 \\
\hline
3 &   1& 2& 4& 5 \\
\hline
4 &   2& 1& 3& 5 \\
\hline
5 &   1& 2& 3& 4 \\
\hline
\end{tabular}
\end{center}
\caption{
Index for vertices of a 4-simplex $v(f,k)$ as a function 
of the index for face (3-simplex) $f$ and the index of vertices of the face $k$.
}
\label{t1}
\end{table}

\begin{figure}
        \includegraphics[height=7cm,width=14cm]{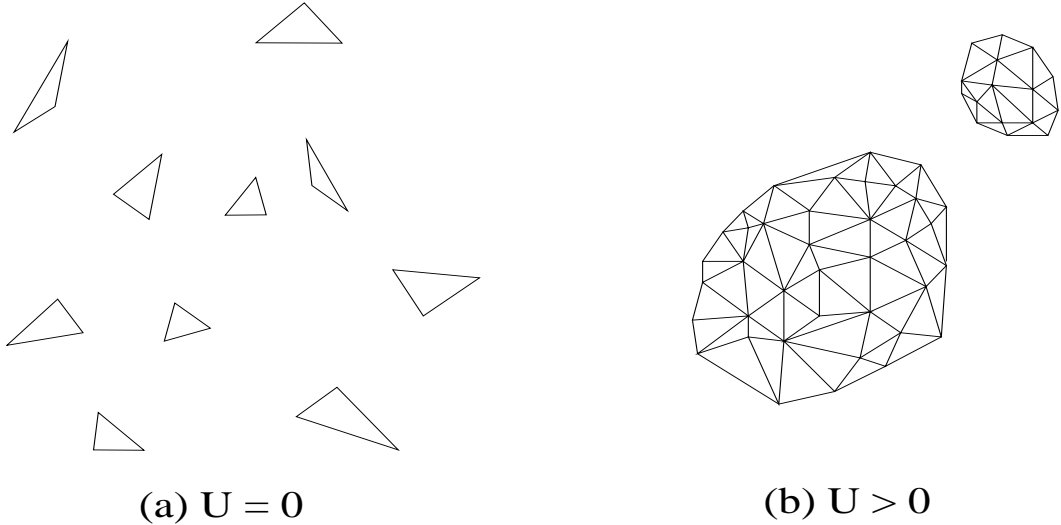}
\caption{
For $U=0$, there is no correlation between simplices in the target space and simplices form a gaseous phase.
For $U > 0$, simplices form closed clusters with no free face.
}
\label{fig:p}
\end{figure}

If $U =0$, there is no correlations and simplices will fluctuate independently in the target space.
The volume of each simplex will be controlled by the `one-body' term, $S_1$.
This is a gas phase of simplices.
This is schematically shown in Fig. \ref{fig:p} (a).
If $U > 0$, there are attractive interactions between faces of simplices
which have opposite orientations.
Each simplex will form a bond with other simplices on the faces.
Since the attractive interaction is singular, simplices will form a bound state 
for any nonzero $U$.
Once a face of a simplex forms a bond with a face of another simplex, 
the face becomes `inert'.
A third simplex can not form a bond on the same face
because the bonded face is made of two faces with opposite orientations
and the interaction between a third simplex is canceled.
In this sense, we can view an open face of a simplex as a charge.
For $U>0$, there is no open face and 
simplices form clusters which 
are oriented and closed without a boundary.
This is illustrated in Fig. \ref{fig:p} (b).
In the following, we will focus on the case with $U>0$.

Once the simplices form clusters, the low energy degrees of freedom are 
the coordinates of the vertices of the clusters in the target space. 
The effective potential can be written as a
sum over linked clusters with different sizes and topologies,
\bqa
\ln Z = \sum_{M,T} z^M  Z_{M,T},
\label{lnZ}
\eqa
where $Z_{M,T}$ is the partition function of $M$ simplices which form a 
closed linked cluster with a topology $T$, and
$z = e^{\mu}$ is the fugacity.
Here, we focus on $Z_{M,T}$ with a fixed topology, i.e. $S^4$, in the limit of $M \rightarrow \infty$.
For a given topology, different ways of creating a cluster of simplices correspond to different ways of triangulating the 4D space of the given topology in the target space.
Then the partition function $Z_{M,T}$ can be written as
\bqa
Z_{M,T} & = & \sum_t \frac{1}{C_{t}} \int \Pi_i \Pi_A d \phi^A_i e^{-\sum_s f[ V_s ]}.
\label{ZMT}
\eqa
Here, $t$ runs over all possible triangulations of the 4D manifold
with fixed topology $T$ using the fixed number of simplices $M$.
$C_{t}$ is the symmetry factor for the triangulation $t$, that is, the
number of permutations of simplices and vertices under which the triangulation remains invariant.
$i$ is an index for vertices of the linked cluster and $\phi_i$ is the coordinate of the i-th vertex in the target space.
For a given triangulation, the coordinates of vertices, $\{ \phi_i \}$ completely characterize the shape of a cluster. 
We can view the cluster as a discrete membrane embedded in the $N$-dimensional target space.

The summation over all possible triangulations in Eq. (\ref{ZMT})
has the similar structure as the one in the Euclidean dynamical triangulation
with an infinite gravitational coupling in which case the crumpled phase with an infinite Hausdorff dimension is stable\cite{EDT}.
If the simplices form an infinitely crumpled cluster, there will be no extended space.
However, the theory in Eq. (\ref{ZMT}) has a crucial difference from the dynamical triangulation.
In Eq. (\ref{ZMT}), the vertices are dynamical variables fluctuating in the target space
while in the dynamical triangulation there is no such dynamical variable except for 
the way simplices are glued together.
The crumpled phase is unlikely in the present theory at least for a large $N$ for the following reason.
In the crumpled phase, a huge number of simplices share a small number of vertices.
This results in only a small number of fluctuating vertices in the $N$-dimensional target space
and the entropy is highly suppressed compared to one in the phase which has a macroscopically large number of vertices.
The entropy contribution is important for a large $N$ and the crumpled phase is unlikely to be stable.
Therefore, we expect that the simplices will form a discrete membrane with a finite Hausdorff dimension $d_H$.
At this stage, we can not exclude the possibility that $d_H$ is larger or smaller than $4$.
Here we assume that $d_H=4$ and 
proceed to show that a four dimensional gravitational theory 
will emerge at low energies under this assumption.
A similar idea for an emergent space and time has been pursued based on a graph model\cite{SMOLIN}.

It is noted that the membrane can superficially look highly `crumpled' in the target space even though it has the Hausdorff dimension $4$.
What is important in determining the Hausdorff dimension is the connectivity of simplices.
If simplices are connected to each other in such a way that it can be unfolded into a 4-dimensional extended membrane, the discrete membrane has the Hausdorff dimension $4$ even though the image of the membrane are highly localized in the target space.
On the other hand, if simplices are highly connected in such a way that a simplex can be reached from any other simplex through a finite number of simplices, the simplices can not be unfolded into an extended membrane and the space does not have a finite Hausdorff dimension.

\begin{figure}
        \includegraphics[height=7cm,width=14cm]{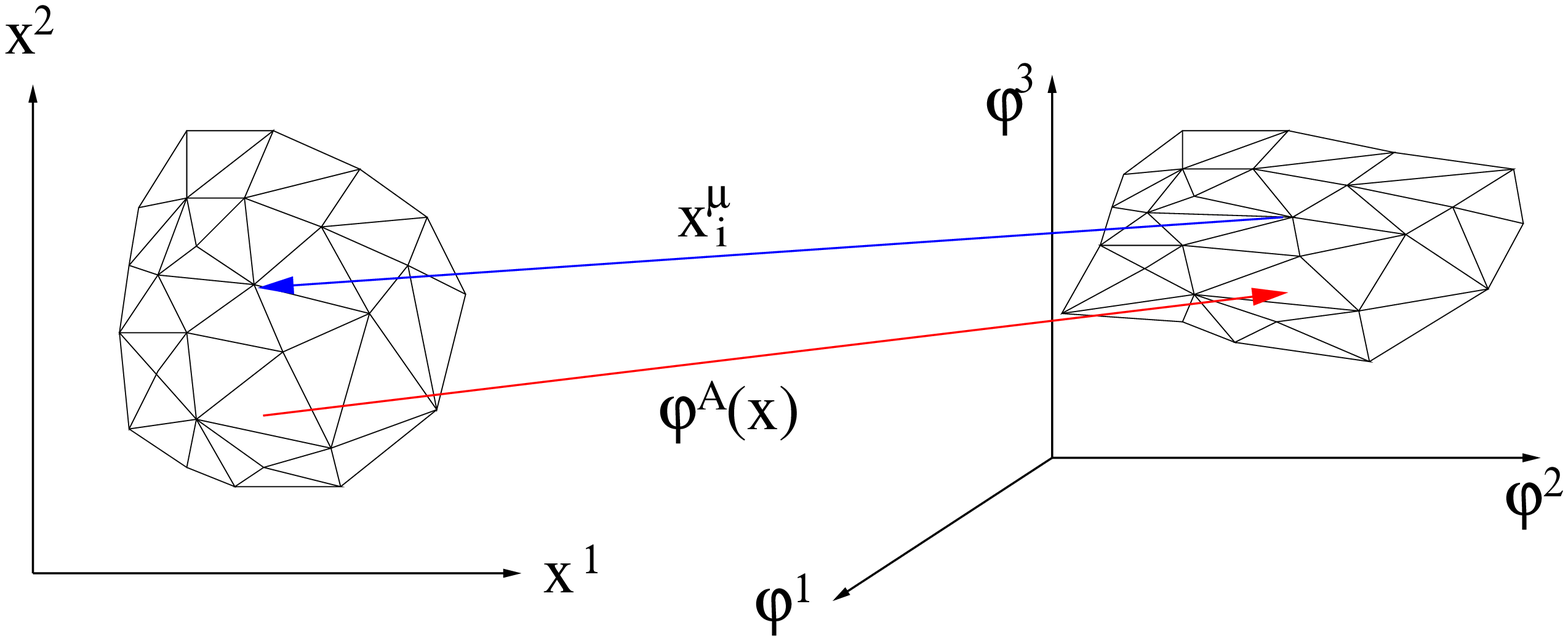}
\caption{
Coordinate space and target space in 2D.
$x^\mu_i$ is a mapping from vertices $i$ to the coordinate space. 
$\phi^A(x)$ is constructed by linearly interpolating the values of $\phi^A_i$
to the interior of each simplex.
}
\label{fig:target}
\end{figure}

Now we introduce a coordinate system for the four dimensional discrete membrane.
We assign a set of coordinates to vertices of the discrete membrane,
$x^\mu_i$ ($\mu=1,2,3,4$) such that orientations of simplices in the coordinate space
is the same as the orientations of the discrete membrane in the target space.
Then we can regard the coordinates of vertices in the target space as boson fields defined on the coordinate space.
Although $\phi^A(x)$ is defined only at the discrete points $x=x_i$,
we can extend the definition to the whole coordinate space by linearly
interpolating the values at the corners of each simplex to the interior of simplex.
There exists a unique linear extension. 
The coordinate space, target space and linear extension of the boson fields
are depicted in Fig. \ref{fig:target} for 2D with $N=3$.
The generalization to the four dimensional case with a larger N is straightforward although it is hard to visualize.
Then one can rewrite the action using $\phi^A(x)$.

Using the extended boson field on the coordinate space, 
we can express the volume of a simplex in the target space as
$
V_s[\phi] = v_s \sqrt{ h_s[\phi] },
$
where $v_s = \int_s d^4 x$ is the volume of simplex $s$ in the coordinate space 
calculated using the flat Euclidean metric
and $h_s[\phi]$ is the determinant of a matrix defined by
$
H_{s,\mu \nu}[\phi] = \frac{ \partial \phi^A}{\partial x^\mu}  \frac{ \partial \phi^A}{\partial x^\nu},
$
where repeated indices are summed.
Note that  $H_{s, \mu \nu}[\phi]$ is constant inside each simplex 
because $\phi^A(x)$ is piecewise linear.
This theory is difficulty to analyze because
the action is highly nonlinear as a function of the fundamental degrees of freedom $\phi^A$.

To obtain a quadratic action for the fundamental field $\phi^A$, we perform a `generalized Hubbard-Stratonovich transformation' for 
Eq. (\ref{p}), that is, to rewrite the partition function 
of $e^{-f( v_s \sqrt{h_s} )}$ using the auxiliary field in Eq.(\ref{f}).
In this transformation, 
$16$ auxiliary fields ${\bf E}_s$ are introduced in each simplex.
Introducing a symmetric matrix ${\bf C}_s$ which satisfies
${\bf H}_s = {\bf C}_s^{2}$\cite{C},
we change the auxiliary variables,
${\bf E}_s   =  ( v_s \Lambda \sqrt{h_s} )^{1/4} {\bf C}_s^{-1} {\bf E}_s^{'}$,
where $\Lambda$ is a constant with dimension $4$
which ensure that ${\bf E}_s^{'}$ has the dimension $0$ as ${\bf E}_s$ does\cite{D}.
In terms of the new variables, the partition function becomes 
\bqa
Z_{M,T} & = & \sum_{t} \frac{1}{C_{t}} \int d \phi^A_i d {\bf E}_s^{'} \Pi_s ( v_s \Lambda)^4  
e^{-S}
\label{zs}
\eqa
with
\bqa
S &  = & 
 \sum_s v_s \sqrt{g_s^{'}} 
\Bigl[  \Lambda + \sum_{n=1}^\infty \alpha_n  \Lambda^{(2-n)/2} {\mbox tr}( {\bf G}_s^{'-1} {\bf H}_s )^n
\Bigr],
\label{p2}
\eqa
where ${\bf G}^{'}_s = {\bf E}^{'}_s {\bf E}^{'T}_s$ and $g_s^{'} = det {\bf G}_s^{'}$.
Redefining the boson fields so that they have the usual scaling dimension $1$,
$
\phi^{A'}_i  =  \Lambda^{1/4} \phi^{A}_i,
$
we obtain the action 
\bqa
S & = &
 \sum_s v_s \sqrt{g_s^{'}} 
 \Bigl[
 \Lambda 
+ \alpha_1 G_s^{' \mu\nu} H_{s, \mu\nu}^{'}   
+ \frac{\alpha_2}{ \Lambda} G_s^{' \mu\nu} H_{s, \nu\lambda}^{'}   G_s^{' \lambda\sigma} H_{s, \sigma\mu}^{'} 
+ ...
\Bigr]  ,
\label{p3}
\eqa
where $G_s^{' \mu\nu}$ is the inverse of $G^{'}_{s, \mu\nu}$ 
and 
$H_{s, \mu \nu}^{'} = \frac{ \partial \phi^{A'}}{\partial x^\mu}  \frac{ \partial \phi^{A'}}{\partial x^\nu}$
written in component form.

Eq. (\ref{p3}) is a discrete action of $N$ scalar fields coupled to gravity with a positive cosmological constant $\Lambda$.
The auxiliary field $E_{\mu a}$ corresponds to the tetrad and $G_{\mu \nu}$, the metric.
The metric field does not have a curvature term and 
the bare gravitational coupling is infinite.
However, the curvature term for the metric field will be dynamically
generated once short distance modes of the matter fields are integrated out.
At low energies, the metric field becomes a propagating mode. 
The $N$ scalar bosons are coupled to the gravitational field with the same strength (the universality of the gravitational coupling).
This is due to the $SO(N)$ rotational symmetry in the target space.
The cosmological constant and other coupling constants will be also renormalized at low energies.

The induced gravitational action becomes proportional to the number of scalar bosons $N$ and 
we expect that non-perturbative effects are not important for a large $N$.
As the gravitational coupling is renormalized to a finite value of the order of $1/N$ at an intermediate energy scale, 
the theory will keep flowing toward the decoupled theory in the low-energy limit because the gravitational coupling is irrelevant in 4D. 
The suppression of the quantum fluctuations of the metric field in the large $N$ limit can be also understood in terms of the world volume action.
With increasing $N$, the dimension of the target space increases.
In the target space with the large dimension, there is little chance
that an edge in the discrete membrane becomes parallel to nearby edges.
As a result, the simplices are more randomly oriented in the target space and the phase space which does not change the volumes of individual simplices are more suppressed.
This will suppress spikes of the membrane in the target space and there is a more chance that a well-defined continuum limit exists for a large N.
For a N which is smaller than a certain critical value, the partition function can develop a singularity for configurations with infinitesimally small world volume in the target space because of the logarithmic singularity in the one-body potential (\ref{f4}).
This will lead to a complete collapse of the 4D membrane to a point in the target space.
Such state can be regarded as a gravitational confinement phase.
As argued before, the confinement phase is unlikely for a large enough $N$.

Here, we examine two possible phases as we tune $\alpha_n$.
If  $\alpha_n > 0$ for all $n$, 
the scalar fields have the usual kinetic energy term 
$\alpha_1 G^{\mu\nu} \frac{ \partial \phi^{A}}{\partial x^\mu}  \frac{ \partial \phi^{A}}{\partial x^\nu}$ 
and they fluctuate in a small region of the target space.
Since the energy increases as $ \frac{ \partial \phi^{A}}{\partial x^\mu}$ increases,
a positive $\alpha_1$ corresponds to a positive tension of the membrane and
the size of simplices in the membrane remains small in the target space.
In this phase, $N$ massless bosons and the dynamical metric field
arise as low energy excitations.
For generic $\Lambda$, the saddle point configuration of the metric is not flat.
Nonetheless, there will be graviton modes propagating in the curved background
if the renormalized curvature is small compared to the underlying cut-off scale.
To achieve this, a fine tuning of microscopic parameters ($\alpha_n$) 
may be necessary.
This phase is a `gravitational Coulomb phase' which is analogous to the Coulomb (deconfinement) phase 
in gauge theories where a massless gauge boson arises as a low-energy excitation.

\begin{figure}
        \includegraphics[height=4cm,width=8cm]{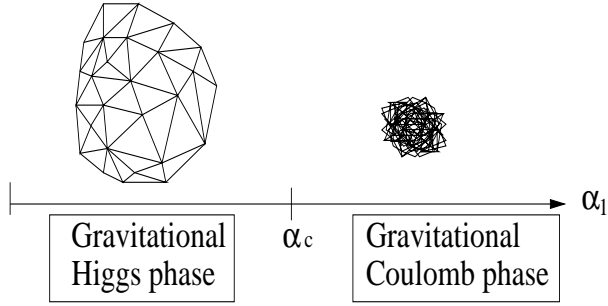}
\caption{
Schematic phase diagram with $U>0$ in the large N limit.
In the `gravitational Higgs phase' ($\alpha_1 < 0$), the membranes are extended in the target space.
In the `gravitational Coulomb phase' ($\alpha_1 > 0$), the membrane fluctuates in a small region of the target space.
}
\label{fig:phase}
\end{figure}

If $\alpha_1 < 0$ and $\alpha_n>0$ for $n>1$, the  kinetic energy of the scalar field has a negative coefficient
and the action is minimized when $ \frac{ \partial \phi^{A}}{\partial x^\mu} \neq 0$.
In the membrane picture, this corresponds to a negative tension.
Every simplex occupies a finite volume in the target space 
and the gross size of the membrane becomes macroscopically large.
Here we assume the simplest case where the extended membrane 
is flat in the target space and 
the direction of the membrane is determined by a spontaneous symmetry breaking.
Suppose that the membrane is extended in the $\{ \phi^1, \phi^2, \phi^3, \phi^4 \}$ directions.
In this phase, we expect that there is no graviton mode
but there exist only massless bosons at low-energy
which correspond to the longitudinal/transverse fluctuations of the membrane.
This is a gravitational analogue of the Higgs phase\cite{ARKANI}.
The schematic phase diagram is shown in Fig. \ref{fig:phase}.

\begin{table}
 \begin{center}
 \begin{tabular}{|c||c|c|}
\hline
& ~~ gauge theory ~~ & ~~ gravity ~~ \\
\hline
\hline 
~fundamental  degrees of freedom ~& XY angles ($\theta^{ab}_i$) & 4-simplices ($\phi^A_{sv}$) \\
\hline
dynamical constraint & $\theta^{ab}_i + \theta^{bc}_i + \theta^{ca}_i = 0$  &  no free face \\
\hline
low energy mode & slave-particle fields ( $\phi^a_i$ ) & 4D discrete membrane ($\phi^A_i$) \\
\hline
Hubbard-Stratonovich field & complex fields ($\eta_{ij} = |\eta_{ij}|e^{ia_{ij}}$) & $4\times4$ real matrices (${\bf E}_s$) \\
\hline
collective excitations & emergent photon ($a_{\mu}$) & emergent graviton ($G_{\mu\nu}$) \\
\hline
low energy excitations & ~gauge boson and matter bosons ~& ~graviton and matter bosons~ \\
\hline
\end{tabular}
\end{center}
\caption{
Comparison between the emergent  gauge theory and the emergent gravity.
}
\label{t2}
\end{table}

\section{Discussion}

In summary, we examined the possibility that a gravitational theory
emerges from a local model of interacting simplices.
If there is a strong interaction between simplices, the simplices form 
discrete membranes in the target space.
For certain one-body potential for the simplices, 
the effective action for the discrete membranes can be transformed into 
a discrete gravitational theory with an infinite gravitational coupling.
Although there is no bare curvature term,
the metric field acquires a dynamics at low energies due to quantum fluctuations of 
the matter fields and becomes a propagating mode through the Sakharov's mechanism.
We interpret the propagating graviton as an emergent collective excitation of the simplices.
The way the gravitational theory emerges from the system of simplices is closely related to the mechanism
that gauge theories emerge in condensed matter systems through slave-particle theories.
We show table \ref{t2} which compares between the emergent gauge theory and the emergent gravity.

Finally, we conclude with some open questions.
First, in this paper we assumed that the simplices form a cluster with Hausdorff dimension $4$ in the strongly interacting phase.
Although it is likely that the cluster does not have an infinite Hausdorff dimension as argued in the paper, this assumption still needs to be checked.
Second, the interaction potential between simplices considered in the paper is the delta function potential in the target space. 
This is a singular interaction and the `critical point' that separates the weakly interacting phase and the strongly interacting phase is $U=0$.
It will be of interest to find a model where the weakly interacting gaseous phase of simplices are stable over a finite range of microscopic parameters.
Third, it remains to be checked whether the present simplex model indeed has a continuum limit 
where the low energy excitations are described by the slowly varying metric fields.
In the future, it is of great interest to perform a numerical simulation on the model to see whether 
there exists a spin-2 massless mode.

\section{Acknowledgment}
I thank Matthew Fisher, Yong Baek Kim and Xiao-Gang Wen for helpful comments.
I am particularly grateful to Joe Polchinski and Steven Carlip for pointing out erroneous statements on diffeomorphism invariance in the earlier version.
This research was supported in part by the National Science Foundation
under Grant No. PHY99-07949 and NSERC.


\begin{thebibliography}{10}
\bibitem{NONLOCAL} If the metric were a propagating mode from the beginning at the high energy cut-off scale, 
the resulting action obtained after integrating out the metric field would be non-local. 
Such a non-local action would have no chance to be realized as a local microscopic system which
has only the matter field as a fundamental degree of freedom.
\bibitem{D} We assign scaling dimension with the reference to the coordinate space,
that is, $[x^\mu] = -1$ and $[\phi^A] = 0$.
\end{thebibliography}
\end{document}